\documentclass[12pt]{article}

\usepackage[totalheight = 23cm, totalwidth = 17cm]{geometry}
\usepackage{amssymb,amsmath,amsfonts,amsbsy,graphicx}

\newcommand{\fnl}{f_{\rm NL}}
\newcommand{\calA}{{\cal A}}
\newcommand{\calL}{{\cal L}}
\newcommand{\calP}{{\cal P}}

\begin{document}

\begin{titlepage}

\begin{center}

\rightline{CERN-PH-TH/2011-096}

\vskip 1.5cm

\Huge{Large non-Gaussianity in non-minimal inflation}

\vskip 1cm

\large{
Jinn-Ouk Gong\footnote{jinn-ouk.gong@cern.ch}
\hspace{0.2cm} and \hspace{0.2cm}
Hyun Min Lee\footnote{hyun.min.lee@cern.ch}
\\
\vspace{0.5cm}
{\em
Theory Division, CERN, CH-1211 Gen\`eve 23, Switzerland
}}

\vskip 0.5cm

\today

\vskip 1.2cm

\end{center}

\begin{abstract}

We consider a simple inflation model with a complex scalar field coupled to gravity non-minimally. Both the modulus and the angular directions of the complex scalar are slowly rolling, leading to two-field inflation. The modulus direction becomes flat due to the non-minimal coupling, and the angular direction becomes a pseudo-Goldstone boson from a small breaking of the global $U(1)$ symmetry. We show that large non-Gaussianity can be produced during slow-roll inflation under a reasonable assumption on the initial condition of the angular direction. This scenario may be realized in particle physics models such as the Standard Model with two Higgs doublets.

\end{abstract}

\end{titlepage}

\setcounter{page}{0}
\newpage
\setcounter{page}{1}

\section{Introduction}

Cosmic inflation~\cite{inflation} has been the cornerstone beyond the standard big bang cosmology. It makes the present observable universe homogeneous and isotropic on large scales and gives rise to the primordial perturbations~\cite{book}. These primordial perturbations, when reentering the horizon after inflation, become the seed for the subsequent structure formation that leads to the temperature anisotropies in the cosmic microwave background and the inhomogeneous distribution of galaxies. They are very well constrained by observations to follow Gaussian statistics and to give nearly scale invariant power spectrum~\cite{Komatsu:2010fb}. Although there have been many proposals for inflation models consistent with observations~\cite{Lyth:1998xn}, it seems hard to distinguish between different models from the current observations.

As we are getting more precise cosmological data from the ongoing and future experiments, e.g. from the Planck satellite~\cite{:2006uk}, we will be able to constrain the models of inflation beyond Gaussianity and power spectrum in next few years. Especially, the possibility of observing primordial non-Gaussianity from inflation has recently drawn a lot of attention from both cosmologists and particle physicists~\cite{nGreviews}. Since most slow-roll inflation models predict negligible non-Gaussianity~\cite{Tanaka:2010km}, the detection or null-detection of non-Gaussianity will be an important discriminator for inflation models.

The inflation models with non-minimal couplings~\cite{nonminimalinf} have been receiving a renewed interest from the recent claim that the Standard Model (SM) Higgs field can support inflation provided that it is non-minimally coupled to gravity~\cite{Bezrukov:2007ep}. The merit of the Higgs inflation consists in the minimality that the inflaton is the Higgs boson in the SM with a single additional parameter, the non-minimal coupling of the Higgs doublet. However, matching to the COBE normalization for the resulting density perturbation, one needs a large non-minimal coupling, given that the Higgs self-coupling appears large to be consistent with the Higgs mass bound at LEP~\cite{lep}. It has been shown that the large non-minimal coupling gives rise to a unitarity problem in the gauge boson scatterings of the SM~\cite{unitarity} so the perturbative expansion breaks down at unitarity scale without adding extra degrees of freedom~\cite{Giudice:2010ka,unitarity2}. Nonetheless, the predictions of the Higgs inflation may be maintained due to the field-dependent unitarity cutoff~\cite{Bezrukov:2010jz} and/or depending on a ultraviolet completion~\cite{Giudice:2010ka,uvcomplete}. Therefore, the Higgs inflation seems to remain as an interesting proposal.

In this paper, we consider a simple two-field inflation model with a complex scalar field. We assume that the non-minimal coupling is not large such that there is no unitarity scale much below the Planck scale. The potential of the modulus of the complex scalar becomes flat at large field value due to the non-minimal coupling. Moreover, a small breaking of the global $U(1)$ symmetry gives rise to a flat potential for the angular mode of the complex scalar too. On the other hand, the slow-roll parameters of the angular mode remain small during inflation such that the slow-roll conditions for the angular mode are always satisfied. At horizon crossing, the vacuum expectation value of the modulus is chosen to obtain the required number of $e$-folds while that of the angular mode is set in such a way that it gives a negligible contribution to the slow-roll parameter $\epsilon$. Very interestingly, we show that a sizable contribution of the angular mode to $\epsilon$ at the end of inflation leads to non-Gaussianity large enough to be detected.

This paper is organized as follows. In Section~\ref{sec:model}, we first present the inflation model with a complex scalar field non-minimally coupled to gravity and show how two-field inflation is possible. Then, in Section~\ref{sec:pert} we analysis the perturbations during inflation using the $\delta{N}$ formalism~\cite{deltaN} and compute the non-linear parameter $\fnl$ as well as the power spectrum $\calP_\zeta$ and the spectral index $n_\zeta$. Finally, in Section~\ref{sec:conc} conclusions are drawn.

\section{Model}
\label{sec:model}

We consider the general action for a complex scalar field $\phi$ with dimension-4 interactions in the Jordan frame,
\begin{align}\label{JF_action}
\frac{\calL_J}{\sqrt{-g_J}} = & \frac{1}{2}R + \left( \xi|\phi|^2 + \frac{1}{2}\alpha\phi^2 + \frac{1}{2}\alpha{\phi^*}^2 \right)R - |\partial_\mu\phi|^2 - V_J(\phi) \, ,
\\
V_J(\phi) = & \lambda|\phi|^4 + \Big(\frac{1}{2}\kappa\phi^3\phi^* +\frac{1}{2}\zeta \phi^4+{\rm h.c.}\Big) \, .
\end{align}
Here we note that $\xi$ and $\lambda$ respect the $U(1)$ global symmetry while $\alpha$, $\kappa$ and $\zeta$ do not. For simplicity, we choose all the dimensionless couplings to be real. Then, the unbounded below constraints give $\lambda+\kappa+\zeta>0$, $\lambda-\kappa+\zeta>0$ and $\sqrt{(\lambda+\zeta)^2-\kappa^2}+\lambda-3\zeta>0$.

Writing the complex field $\phi$ in terms of two real fields $h$ and $\chi$ as
\begin{equation}
\phi = \frac{h}{\sqrt{2}}e^{i\chi/f} \, ,
\end{equation}
the original action (\ref{JF_action}) becomes
\begin{align}
\frac{\calL_J}{\sqrt{-g_J}} = & \frac{1}{2}R + \left[ \frac{1}{2}\xi h^2 + \frac{1}{2}\alpha h^2\cos\left( \frac{2\chi}{f} \right) \right]R - \frac{1}{2}(\partial_\mu h)^2 - \frac{h^2}{2f^2}(\partial_\mu\chi)^2
\nonumber\\
&- \frac{1}{4}\lambda h^4 - \frac{1}{4}\kappa h^4\cos\left( \frac{2\chi}{f} \right) - \frac{1}{4}\zeta h^4\cos\left( \frac{4\chi}{f} \right) \, .
\end{align}
Changing to the Einstein frame by a conformal transformation with
\begin{equation}
\Omega^2 \equiv 1 + \xi h^2 +\alpha h^2\cos\left( \frac{2\chi}{f} \right) \, ,
\end{equation}
we have
\begin{align}\label{EF_action}
\frac{\calL_E}{\sqrt{-g_E}} = & \frac{1}{2}R - \frac{1}{2\Omega^2} \left\{ 1 + \frac{6h^2}{\Omega^2} \left[ \xi + \alpha\cos\left( \frac{2\chi}{f} \right) \right]^2 \right\} (\partial_\mu h)^2 - \frac{h^2}{2f^2\Omega^2} \left[ 1 + \frac{6\alpha^2h^2}{\Omega^2}\sin^2\left( \frac{2\chi}{f} \right) \right] (\partial_\mu\chi)^2
\nonumber\\
& + \frac{6\alpha h^3}{f\Omega^4} \sin\left( \frac{2\chi}{f} \right) \left[ \xi + \alpha\cos\left( \frac{2\chi}{f} \right) \right] \partial^\mu h\partial_\mu\chi - \frac{h^4}{4\Omega^4} \left[ \lambda + \kappa\cos\left( \frac{2\chi}{f} \right)+\zeta \cos\left( \frac{4\chi}{f} \right)\right] \, .
\end{align}
Thus, for a non-zero $\alpha$, there is a kinetic mixing between the modulus and angular directions. For large values of the modulus, the potential becomes flat along the modulus as in Higgs inflation. Meanwhile, the potential for the angular direction may be too steep if the symmetry breaking terms proportional to $\alpha$, $\kappa$ and $\zeta$ are of order unity. To see this more clearly, we take the limit $h\gg1/\sqrt{\xi}$. Then, (\ref{EF_action}) becomes
\begin{align}
\frac{{\cal L}_E }{\sqrt{-g_E}} \approx & \frac{1}{2}R - \frac{3}{h^2}(\partial_\mu h)^2
-\frac{1}{2\xi f^2} \left[ 1+\frac{3\alpha^2}{\xi}\sin^2\left(\frac{2\chi}{f}\right) \right](\partial_\mu\chi)^2 + \frac{6\alpha}{\sqrt{\xi} h}\sin^2 \left(\frac{2\chi}{f}\right) \partial^\mu h\partial_\mu\chi
\nonumber\\
& - \frac{1}{4}\,\frac{\lambda+\kappa \cos(2\chi/f)+\zeta \cos(4\chi/f)}{\left[\xi+\alpha \cos(2\chi/f) \right]^2} \, .
\end{align}
Then, in the potential, the first term proportional to $\lambda$ provides a vacuum energy for inflation while the pseudo-scalar $\chi$ dependent terms can satisfy the slow-roll conditions for $\kappa$, $\zeta\ll \lambda$ and $\alpha\ll \xi$. The small $U(1)$ breaking terms are technically natural in the sense that for equally small $U(1)$ breaking parameters, the loop corrections to them are under control. In this case, $\chi$ can be another inflaton component. The potential for $\chi$ is of a very similar form to the inflaton potential for a codimension-two brane moving in a warped background in six dimensions~\cite{cod2brane}. In the brane inflation case, inflation is assumed to end when the moving brane hits the background brane\footnote{But this is not always the case: see e.g. Ref.~\cite{branetip}.}. On the other hand, in our case, the modulus is the main inflaton component,  which finishes inflation when the slow-roll condition is not satisfied any more.

From now on, for simplicity, we consider the case with $\alpha=0$ so that the kinetic mixing is absent. Also we work in the Einstein frame and drop the subscript $E$. The results remain valid for the more general case with non-zero $\alpha$ as far as $|\alpha|\ll \sqrt{\xi}$. In this case, (\ref{EF_action}) takes a much simpler form,
\begin{equation}
\frac{\calL}{\sqrt{-g}} \approx \frac{1}{2}R - \frac{1}{2\Omega^2} \left( 1 + \frac{6\xi^2h^2}{\Omega^2} \right) (\partial_\mu h)^2 - \frac{h^2}{2f^2\Omega^2} (\partial_\mu\chi)^2
- \frac{h^4}{4\Omega^4} \left[ \lambda + \kappa\cos\left( \frac{2\chi}{f}\right)+\zeta \cos\left( \frac{4\chi}{f}\right) \right] \, .
\end{equation}
We can find the canonical radial field $\varphi$ from
\begin{equation}
\frac{d\varphi}{dh} = \frac{1}{\Omega} \sqrt{ 1 + \frac{6\xi^2h^2}{\Omega^2} }
\end{equation}
as
\begin{equation}
\varphi = \sqrt{1+6\xi^{-1}} \sinh^{-1} \left[ \sqrt{\xi(1+6\xi)}h \right] - \sqrt{2}\tanh^{-1} \left[ \frac{\sqrt{6}\xi h}{\sqrt{1+\xi(1+6\xi)}h^2} \right] \, .
\end{equation}

Let us consider two different regime of $h$. First, for $0<h\ll1/\xi$, $\varphi \approx h/\Omega$ and
\begin{equation}
V(\varphi,\chi) \approx \frac{1}{4}\varphi^4 \left[ \lambda + \kappa\cos\left( \frac{2\chi}{f} \right) +\zeta \cos\left( \frac{4\chi}{f}\right)\right] \, .
\end{equation}
Meanwhile, for $h\gg1/\xi$, we have a non-trivial relation
\begin{equation}
\frac{d\varphi}{dh} \approx \frac{\sqrt{6}\xi h}{1+\xi h^2} \, ,
\end{equation}
from which we obtain
\begin{equation}
\varphi \approx \frac{\sqrt{6}}{2}\log \left( 1 + \xi h^2 \right) \, .
\end{equation}
Then, in this limit, the potential is approximated by
\begin{equation}
V(\varphi,\chi) \approx \frac{\lambda}{4\xi^2} \left( 1 - e^{-2\varphi/\sqrt{6}} \right)^2 \left[ 1 + \delta_1\cos \left( \frac{2\chi}{f} \right) + \delta_2 \cos\left( \frac{4\chi}{f}\right) \right] \, ,
\end{equation}
where $\delta_1\equiv\kappa/\lambda$ and $\delta_2\equiv\zeta/\lambda$. Therefore, we have included a small potential term for the modulus $\varphi$ as well so that both the modulus and the angular directions of the complex scalar are slowly rolling. In order to canonically normalize the kinetic term for the angular direction $\chi$ during inflation, we need to choose $f=1/\sqrt{\xi}$, viz. the decay constant of the would-be Goldstone boson is $1/\sqrt{\xi}$. This is comparable to a similar observation in Higgs inflation that the gauge boson mass during inflation is of $\mathcal{O}(1/\sqrt{\xi})$, which is identified with the ultraviolet cutoff during inflation~\cite{Bezrukov:2010jz}.

Focusing on the regime with $h\gg 1/\xi$, we can write the Lagrangian as
\begin{equation}\label{EF_action2}
\frac{\calL}{\sqrt{-g}} \approx \frac{1}{2}R - \frac{1}{2}(\partial_\mu\varphi)^2 - \frac{1}{2} e^{2b(\varphi)}(\partial_\mu\chi)^2 - W(\varphi,\chi)\, .
\end{equation}
This is a particularly simple Lagrangian~\cite{GarciaBellido:1995qq} which has been studied in models motivated from string theory: it includes a non-canonical kinetic term with
\begin{equation}
e^{2b(\varphi)} \equiv 1 - e^{-2\varphi/\sqrt{6}} \, ,
\end{equation}
and a potential of product form $W(\varphi,\chi) = U(\varphi)V(\chi)$ with
\begin{align}
\label{potentialU}
U(\varphi) = & \frac{\lambda}{4\xi^2} \left( 1 - e^{-2\varphi/\sqrt{6}} \right)^2 \, ,
\\
\label{potentialV}
V(\chi) = & 1 +\delta_1  \cos \left( 2\sqrt{\xi}\chi \right)+\delta_2 \cos\left( 4\sqrt{\xi}\chi \right) \, .
\end{align}
As will be shown later, the above product form of the potential makes it particularly easy to study two-field inflation, not only for the background evolution but also for the explicit computations of the perturbations using the $\delta N$ formalism.

\subsection{Slow-roll parameters}

Since we are interested in the slow-roll phase of inflation, we can write the slow-roll parameters based on the potential derivatives. The first slow-roll parameter $\epsilon$ can be written as
\begin{equation}
\epsilon = \frac{1}{2} \left( \frac{U'}{U} \right)^2 + \frac{e^{-2b}}{2} \left( \frac{V'}{V} \right)^2 \equiv \epsilon^\varphi + \epsilon^\chi \, ,
\end{equation}
where a prime on $U$ and $V$ denotes a derivative with respect to $\varphi$ and $\chi$, respectively. We can also define the second slow-roll parameter $\eta$ as
\begin{align}
\eta^{\varphi\varphi} \equiv & \frac{U''}{U} \, ,
\\
\eta^{\chi\chi} \equiv & e^{-2b}\frac{V''}{V} \, .
\end{align}
Likewise, let us define
\begin{align}
\epsilon^b \equiv & 8{b'}^2 \, ,
\\
\eta^b \equiv & 16b'' \, ,
\end{align}
where a prime on $b$ denotes a derivative with respect to $\varphi$.

With the potential given by (\ref{potentialU}) and (\ref{potentialV}), these slow-roll parameters are given by
\begin{align}
\label{epsilonvarphi}
\epsilon^\varphi = & \frac{4}{3}\frac{e^{-4\varphi/\sqrt{6}}}{\left( 1 - e^{-2\varphi/\sqrt{6}} \right)^2} \, ,
\\
\label{epsilonchi}
\epsilon^\chi = & \frac{2\xi}{1-e^{-2\varphi/\sqrt{6}}}\, \frac{\left[\delta_1\sin\left(2\sqrt{\xi}\chi\right)+2\delta_2\sin^2\left(4\sqrt{\xi}\chi\right)\right]^2}{\left[ 1 + \delta_1\cos\left(2\sqrt{\xi}\chi\right) +\delta_2\cos\left(4\sqrt{\xi}\chi\right)\right]^2} \, ,
\\
\label{etavarphi}
\eta^{\varphi\varphi} = & -\frac{4}{3}e^{-2\varphi/\sqrt{6}} \frac{1-2e^{-2\varphi/\sqrt{6}}}{\left( 1 - e^{-2\varphi/\sqrt{6}} \right)^2} \, ,
\\
\label{etachi}
\eta^{\chi\chi} = & -\frac{4\xi}{1 - e^{-2\varphi/\sqrt{6}}} \,\frac{\delta_1\cos\left(2\sqrt{\xi}\chi\right)+4\delta_2\cos\left(4\sqrt{\xi}\chi\right)}{1 + \delta_1\cos\left(2\sqrt{\xi}\chi\right)+\delta_2\cos\left(4\sqrt{\xi}\chi\right)} \, ,
\end{align}
and
\begin{align}
\epsilon^b = & \epsilon^\varphi \, ,
\\
\eta^b = & -4e^{2\varphi/\sqrt{6}}\epsilon^\varphi \, .
\end{align}
Thus, we find the relation between the slow-roll parameters for $\varphi$ as $\epsilon^\varphi\approx 3\left(\eta^{\varphi\varphi}\right)^2/4$. For $\delta_2=0$, we note that $\epsilon^\chi$ is written in terms of $\eta^{\chi\chi}$ as
\begin{equation}
\epsilon^\chi=\frac{2\xi\delta_1^2}{1 - e^{-2\varphi/\sqrt{6}}} \left\{\frac{1}{\left[ 1+\delta_1\cos\left(2\sqrt{\xi}\chi\right)\right]^2} - \left( 1 - e^{-2\varphi/\sqrt{6}} \right)^2 \left( \frac{\eta^{\chi\chi}}{4\xi\delta_1} \right)^2 \right\} \, .
\label{epetachi}
\end{equation}
Moreover, $\eta^{\chi\chi}$ has an upper bound as
\begin{equation}
\left|\eta^{\chi\chi}_{\rm max}\right| \approx 4\xi|\delta_1| \, .
\end{equation}
In this case, when $\left|\eta^{\chi\chi}\right|$ reaches the maximum value $|\eta^{\chi\chi}_{\rm max}|$, $\epsilon^\chi$ is minimized. For $\delta_2\neq 0$, when $\epsilon^\chi$ is minimized, $|\eta^{\chi\chi}|$ does not have to reach $|\eta^{\chi\chi}_{\rm max}|$ but it is not very different from it.

Slow-roll inflation occurs for $\epsilon\ll1$. This corresponds to $e^{-4\varphi/\sqrt{6}}\ll1$ from (\ref{epsilonvarphi}) and $\xi\delta^2_1$, $\xi\delta_2^2\ll1$ from (\ref{epsilonchi}). From the slow-roll condition for $\epsilon^\varphi$ we find $|\eta^{\varphi\varphi}|\ll1$ automatically. On the other hand, the slow-roll condition on $\eta^{\chi\chi}$ requires $|\delta_1|$, $|\delta_2|\ll1$ for $\xi\gtrsim1$, so the $\chi$ field gives a small contribution to the vacuum energy and inflation is driven dominantly by the radial mode $\varphi$. Then, we can also satisfy $|\epsilon^{\chi\chi}|\ll1$. We note that the slow-roll parameters for $\chi$ can be small in the entire field space. Henceforth, for simplicity, we focus on the case $\delta_1\equiv\delta\neq0$ and $\delta_2=0$. A non-zero but small $\delta_2$ would not change the results very much.

\subsection{Number of $e$-folds}

Slow-roll inflation ends when $\epsilon=1$. Since the slow-roll parameter for $\chi$ remains small, the slow-roll condition is violated {\em mainly} by $\varphi$ when $e^{-2\varphi_e/\sqrt{6}} \approx 0.464$, with the subscript $e$ denotes the end of slow-roll inflation. Here, we refer the readers to later sections for what we here mean by ``mainly''. This gives $\varphi_e \approx 0.940$. This shows that we always have $e^{-2\varphi/\sqrt{6}}\ll 0.464$, which is very convenient to compute the number of $e$-folds $N$ and to determine the other field values $\varphi_\star$, $\chi_\star$ and $\chi_e$, where $\star$ denotes the moment when the scale of our interest exits the horizon.

For a product potential, the number of $e$-folds $N$ is found to be~\cite{Choi:2007su}
\begin{equation}\label{efold0}
N = \int_e^\star \frac{U}{U'} d\varphi = \int_e^*\, e^{2b}\frac{V}{V'} \, d\chi \, .
\end{equation}
Thus, from the first equality we obtain
\begin{equation}\label{efold}
N = \frac{3}{4} \left[ e^{2\varphi_\star/\sqrt{6}} - e^{2\varphi_e/\sqrt{6}} - \frac{2}{\sqrt{6}} \left( \varphi_\star - \varphi_e \right) \right] \, .
\end{equation}
For $N = 60$, we need $e^{2\varphi_\star/\sqrt{6}} \approx 80.5$. Consequently, we can uniquely determine $\varphi_\star \approx 5.37$. From the second equality in (\ref{efold0}) the evolution of $\chi$ during slow-roll inflation is governed by
\begin{equation}
dN = e^{2b}\frac{V}{V'}d\chi = \left( 1 - e^{-2\varphi/\sqrt{6}} \right) \frac{1+\delta\cos\left(2\sqrt{\xi}\chi\right)}{2\sqrt{\xi}\delta\sin\left(2\sqrt{\xi}\chi\right)} d\chi \, .
\end{equation}
Neglecting $e^{-2\varphi/\sqrt{6}}$ which is subdominant, we can analytically integrate this expression to find
\begin{equation}
N \approx \frac{1}{4\xi\delta} \left[ \log\left| \frac{\tan\left(\sqrt{\xi}\chi_e\right)}{\tan\left(\sqrt{\xi}\chi_\star\right)} \right| + \delta\log\left| \frac{\sin\left(2\sqrt{\xi}\chi_e\right)}{\sin\left(2\sqrt{\xi}\chi_\star\right)} \right| \right] \, .
\end{equation}
Thus, ignoring the logarithm multiplied by $\delta$, we obtain
\begin{equation}
\left|\tan\left(\sqrt{\xi}\chi_e\right)\right| \approx \left|\tan\left(\sqrt{\xi}\chi_\star\right)\right| e^{4\xi\delta N} \, .\label{chievolution}
\end{equation}
Thus, for a given $\chi_\star$ which may be arbitrary, we can determine $\chi_e$ uniquely.

Also note that from $\varphi_\star$ and $\varphi_e$,
\begin{align}
e^{2b_\star} = & 1 - e^{-2\varphi_\star/\sqrt{6}} \approx 0.987 \, ,
\\
e^{2b_e} = & 1 - e^{-2\varphi_e/\sqrt{6}} \approx 0.536 \, ,
\end{align}
so we have
\begin{equation}\label{b_difference}
2b_e - 2b_\star \approx -0.611 \, .
\end{equation}
We will use this number in the subsequent sections.

\section{Perturbations during slow-roll inflation}
\label{sec:pert}

In this section, we calculate the primordial perturbations generated during slow-roll inflation using the $\delta{N}$ formalism~\cite{deltaN}. Because $\delta N$ is conformally invariant~\cite{conformalinv}, we may use the $\delta{N}$ formalism safely to compute the primordial perturbations. Note that there do exist isocurvature perturbations, and we must follow the post-inflationary evolution e.g. reheating to see how they contribute to $\Delta{T}/T$ and the matter density field. But it is beyond the scope of the present work so we do not consider this topic here.

\subsection{Linear perturbation}

First, let us consider the power spectrum and the spectral index. We can straightforwardly find the first derivatives of $N$ with respect to the initial field values $\varphi_\star$ and $\chi_\star$ as~\cite{GarciaBellido:1995qq}
\begin{align}
\label{Nphi}
\frac{\partial{N}}{\partial\varphi_\star} = & \frac{s^\varphi_\star}{\sqrt{2\epsilon^\varphi_\star}} \left( 1 - e^{2b_e-2b_\star} \frac{\epsilon^\chi_e}{\epsilon_e} \right) \, ,
\\
\label{Nchi}
\frac{\partial{N}}{\partial\chi_\star} = & \frac{s^\chi_\star}{\sqrt{2\epsilon^\chi_\star}} e^{2b_e-b_\star} \frac{\epsilon^\chi_e}{\epsilon_e} \, ,
\end{align}
where $s^i$ denotes the sign of the potential derivative, i.e.
\begin{equation}
s^i = \left\{
 \begin{split}
  & +1 \qquad \text{if} \qquad V_i > 0
  \\
  & -1 \qquad \text{if} \qquad V_i < 0
 \end{split}
\right. \, .
\end{equation}
This is because, while $\epsilon$ is positive definite, the derivative of potential may be either positive or negative. Then, we can simply substitute (\ref{Nphi}) and (\ref{Nchi}) into the $\delta{N}$ formulae
\begin{align}
\calP_\zeta = & \left( \frac{H}{2\pi} \right)^2 \gamma^{ab}N_aN_b \, ,
\\
n_\zeta - 1 = & -2\epsilon - \frac{2}{N_aN^a} \left( 1 + \frac{R^{abcd}}{3}N_aN_d\frac{V_bV_c}{V^2} - N_aN_b\frac{V^{;ab}}{V} \right) \, ,
\end{align}
where the right hand sides are evaluated at $\star$. Here, $\gamma_{ab}$ is the field space metric with the indices $a$ and $b$ being $\varphi$ and $\chi$, $N_a \equiv \partial{N}/\partial\phi^a$ and the same for $V_a$, $R_{abcd}$ is the Riemann curvature tensor of the field space, and a semicolon denotes a covariant derivative with respect to the field space metric.

\begin{figure}[t]
 \begin{center}
  \includegraphics[width=10cm]{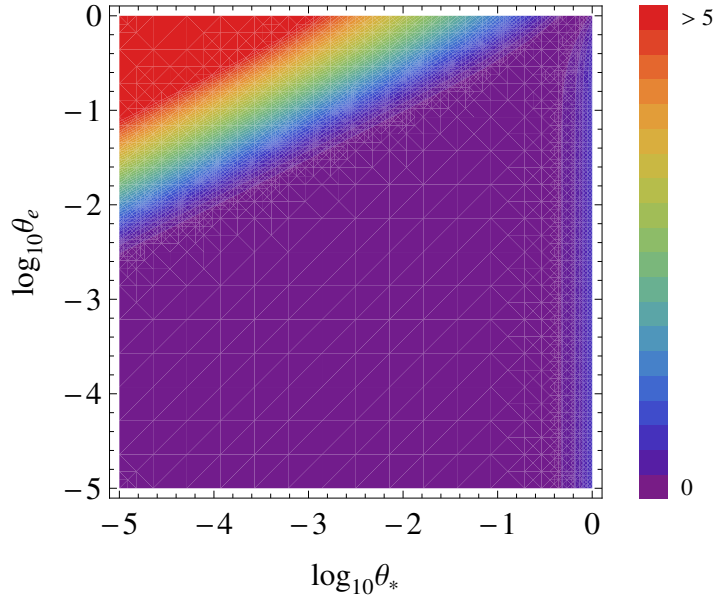}
 \end{center}
 \vspace{-2em}
 \caption{Logarithmic plot of the boost factor multiplied to the single field power spectrum $\left[H_\star/(2\pi)\right]^2/(2\epsilon_\star)$ in (\ref{powerspectrum}) as a function of $\theta_\star$ and $\theta_e$. For a sizeable value of $\theta_e$, this factor may be as large as $10^5$ or even larger. For a large value of $\fnl$, however, we may have a moderate enhancement of $\mathcal{O}(1) \sim \mathcal{O}(10)$: see Figure~\ref{fig:fNLfactor} and Table~\ref{table:numbers}.}
 \label{fig:Pfactor}
\end{figure}

We can conveniently parametrize these quantities from the fact that $\epsilon = \epsilon^\varphi + \epsilon^\chi$. We can introduce a dimensionless angle $\theta$ in such a way that~\cite{Byrnes:2008wi}
\begin{align}
\label{angleparametrization1}
\cos^2\theta \equiv & \frac{\epsilon^\varphi}{\epsilon} \, ,
\\
\label{angleparametrization2}
\sin^2\theta \equiv & \frac{\epsilon^\chi}{\epsilon} \, .
\end{align}
We can see that $\theta$ parametrizes by which field the slow-roll condition is dominated. Further, the difference between $b_e$ and $b_\star$ can be fixed from the background as we have seen in (\ref{b_difference}), so let us write
\begin{equation}
2b_e - 2b_\star \equiv X \, .
\end{equation}
Then, we can straightforwardly find
\begin{align}
\label{powerspectrum}
\calP_\zeta = & \left( \frac{H_\star}{2\pi} \right)^2 \frac{1}{2\epsilon_\star} e^{2X}\frac{\cos^4\theta_e}{\sin^2\theta_\star} \left( \calA^2\tan^2\theta_\star + \tan^4\theta_e \right) \, ,
\\
\label{spectralindex}
n_\zeta - 1 = & -2\epsilon_\star - 4e^{-2X}\frac{\sin^2\theta_\star}{\cos^4\theta_e \left( \calA^2\tan^2\theta_\star + \tan^4\theta_e \right)}\epsilon_\star
\nonumber\\
& + \frac{\cos^2\theta_\star}{12} \frac{\left( \calA\tan^2\theta_\star - \tan^2\theta_e \right)^2}{\calA^2\tan^2\theta_\star + \tan^4\theta_e} \left( \eta^b_\star + 2\epsilon^b_\star \right) \epsilon_\star + \frac{8\calA\sin^2\theta_\star\tan^2\theta_e}{\calA^2\tan^2\theta_\star + \tan^4\theta_e}\epsilon_\star
\nonumber\\
& - \cos^2\theta_\star\tan^2\theta_e \frac{2\calA\tan^2\theta_\star - \tan^2\theta_e}{\calA^2\tan^2\theta_\star + \tan^4\theta_e}\epsilon_\star + \frac{2\left( \calA^2\tan^2\theta_\star\eta^{\varphi\varphi}_\star + \tan^4\theta_e\eta^{\chi\chi}_\star \right)}{\calA^2\tan^2\theta_\star + \tan^4\theta_e} \, ,
\end{align}
where
\begin{equation}
\calA \equiv e^{-X} \left[ 1 + \left( 1-e^X \right) \tan^2\theta_e \right] \, .
\end{equation}
As we can see, the single field result of $\calP_\zeta$ is multiplied by a boost factor that depends on the values of $X$, $\theta_\star$ and $\theta_e$. We show the form of this factor as a function of $\theta_\star$ and $\theta_e$ with a fixed $X$ in Figure~\ref{fig:Pfactor}. Since this factor is always greater than 1, with the observed value of $\calP_\zeta\approx2.5\times10^{-9}$ we may have a considerably lower value of $H_\star$ than the usual single field inflation models. Conversely, this factor would suppress the tensor-to-scalar ratio to an unobservable level. In our model, the Hubble scale during inflation is given by $H_\star^2 \approx \lambda/(12\xi^2)$. When the boost factor in the power spectrum is of order unity, we need $\sqrt{\lambda}\sim10^{-5}\xi$. Therefore, for $\xi\gtrsim 1$, the self-coupling $\lambda$ must be very small. This is different from the usual Higgs inflation.

\subsection{Non-Gaussianity}

By extending the $\delta{N}$ formalism to second order, we can compute the non-Gaussianity generated during slow-roll inflation on super-horizon scales. This corresponds to so-called the local type non-Gaussianity, and is conveniently parametrized by the non-linear parameter $\fnl$~\cite{Komatsu:2001rj}, which is constrained to be $-10 < \fnl < 74$ at $2\sigma$ level~\cite{Komatsu:2010fb}. By taking another derivatives of (\ref{Nphi}) and (\ref{Nchi}), we can obtain~\cite{Choi:2007su}
\begin{align}
\label{Nphiphi}
\frac{\partial^2N}{\partial\varphi_\star^2} = & \left( 1 - \frac{\eta^{\varphi\varphi}_\star}{2\epsilon^\varphi_\star} \right) \left( 1 - e^{2b_e-2b_\star}\frac{\epsilon^\chi_e}{\epsilon_e} \right) + \frac{s^b_\star s^\varphi_\star}{2} \sqrt{\frac{\epsilon^b_\star}{\epsilon^\varphi_\star}} e^{2b_e-2b_\star} \frac{\epsilon^\chi_e}{\epsilon_e} + \frac{e^{4b_e-4b_\star}}{\epsilon^\varphi_\star} {\cal C} \, ,
\\
\label{Nchichi}
\frac{\partial^2N}{\partial\chi_\star^2} = & e^{2b_e} \left( 1 - \frac{\eta^{\chi\chi}_\star}{2\epsilon^\chi_\star} \right) \frac{\epsilon^\chi_e}{\epsilon_e} + \frac{e^{4b_e-2b_\star}}{\epsilon^\chi_\star} {\cal C} \, ,
\\
\label{Nphichi}
\frac{\partial^2N}{\partial\varphi_\star\partial\chi_\star} = & -\frac{s^\varphi_\star s^\chi_\star}{\sqrt{\epsilon^\varphi_\star\epsilon^\chi_\star}} e^{4b_e-3b_\star} {\cal C} \, ,
\end{align}
with $s^b = +1(-1)$ if $b'>(<)0$. Here, $\cal C$ is given by
\begin{equation}
{\cal C} \equiv \frac{\epsilon^\varphi_e\epsilon^\chi_e}{\epsilon_e^2} \left( \eta^{ss}_e - 4\frac{\epsilon^\varphi_e\epsilon^\chi_e}{\epsilon_e} - \frac{s^\varphi_e s^b_e}{2} \sqrt{\frac{\epsilon^b_e}{\epsilon^\varphi_e}} \frac{{\epsilon^\chi_e}^2}{\epsilon_e} \right) \, ,
\end{equation}
where
\begin{equation}
\eta^{ss} \equiv \frac{\epsilon^\varphi\eta^{\chi\chi} + \epsilon^\chi\eta^{\varphi\varphi}}{\epsilon} \, .
\end{equation}

Then, we can compute using the $\delta{N}$ formalism the scale independent part\footnote{We do have a scale dependent part of $\fnl$, which comes from the intrinsic non-Gaussianity in the field fluctuations. But it is always much smaller than 1~\cite{Vernizzi:2006ve} so we do not consider it here.} of $\fnl$ as
\begin{equation}
\frac{6}{5}\fnl = \frac{\gamma^{ac}\gamma^{bd}N_{ab}N_cN_d}{\left(\gamma^{kl}N_kN_l\right)^2} = \frac{1}{\calP_\zeta^2} \left( \frac{H}{2\pi} \right)^4 \left( N_{\varphi\varphi}N_\varphi^2 + 2e^{-2b}N_{\varphi\chi}N_\varphi N_\chi + e^{-4b}N_{\chi\chi}N_\chi^2 \right) \, ,
\end{equation}
where the right hand side is evaluated at $\star$. Using the dimensionless angle $\theta$ introduced in (\ref{angleparametrization1}) and (\ref{angleparametrization2}), we can write $\fnl$ as
\begin{align}\label{nG}
\frac{6}{5}\fnl = & \frac{e^{-X}}{\left( \calA^2\tan^2\theta_\star + \tan^4\theta_e \right)^2} \left( -\frac{\calA^3\tan^4\theta_\star}{\cos^2\theta_e}\eta^{\varphi\varphi}_\star - \frac{\tan^6\theta_e}{\cos^2\theta_e}\eta^{\chi\chi}_\star \right)
\nonumber\\
& + 2e^{-X}\frac{\sin^2\theta_\star}{\cos^2\theta_e} \frac{\calA^3\tan^2\theta_\star + \tan^6\theta_e}{\left( \calA^2\tan^2\theta_\star + \tan^4\theta_e \right)^2}\epsilon_\star + e^{-X} \frac{\sin^2\theta_\star}{\cos^2\theta_e} \frac{\calA^2\tan^2\theta_e\tan^2\theta_\star}{\left( \calA^2\tan^2\theta_\star + \tan^4\theta_e \right)^2}\epsilon_\star
\nonumber\\
& + 2\tan^2\theta_e \frac{\left( \calA\tan^2\theta_\star - \tan^2\theta_e \right)^2}{\left( \calA^2\tan^2\theta_\star + \tan^4\theta_e \right)^2} \left\{ \eta^{\chi\chi}_e\cos^2\theta_e + \sin^2\theta_e \left[ \eta^{\varphi\varphi}_e - \epsilon_e \left( 4\cos^2\theta_e + \frac{1}{2}\sin^2\theta_e \right) \right] \right\} \, .
\end{align}
One can easily verify that among various functions multiplied by the slow-roll parameters, the terms proportional to $\eta^{\chi\chi}$ allow a huge enhancement at small $\theta_\star$ and $\theta_e$. There are two possibilities for large non-Gaussianity\footnote{The two conditions are closely related to those in Ref.~\cite{Byrnes:2008wi}.}:
\begin{enumerate}
\item $\calA^2\tan^2\theta_\star \lesssim \tan^4\theta_e$ and $\tan^2\theta_\star \lesssim \tan^2\theta_e$. In this case, however, the second term of (\ref{spectralindex}) may diverge as $\sin^2\theta_\star/\sin^4\theta_e$ and we need more care.
\item $\calA^2\tan^2\theta_\star \gtrsim \tan^4\theta_e$ and $\tan^2\theta_\star \lesssim \tan^2\theta_e$. In this case, $\eta$'s give dominant contributions to $n_\zeta$. Therefore we focus on this case for the discussion below.
\end{enumerate}
Note that for $\calA^2\tan^2\theta_\star \gtrsim \tan^4\theta_e$ and $\tan^2\theta_\star \gtrsim \tan^2\theta_e$ we have a very small value of $\fnl$, consistent with current observations though, and we do not consider this case. For $\calA^2\tan^2\theta_\star \lesssim \tan^4\theta_e$ and $\tan^2\theta_\star \lesssim \tan^2\theta_e$, $\fnl$ could be large. But this is not possible because $\epsilon^\chi\gtrsim \epsilon^\varphi$ always, which means inflation occurs mostly along $\chi$ direction.

\begin{figure}[t]
 \begin{center}
  \includegraphics[width=10cm]{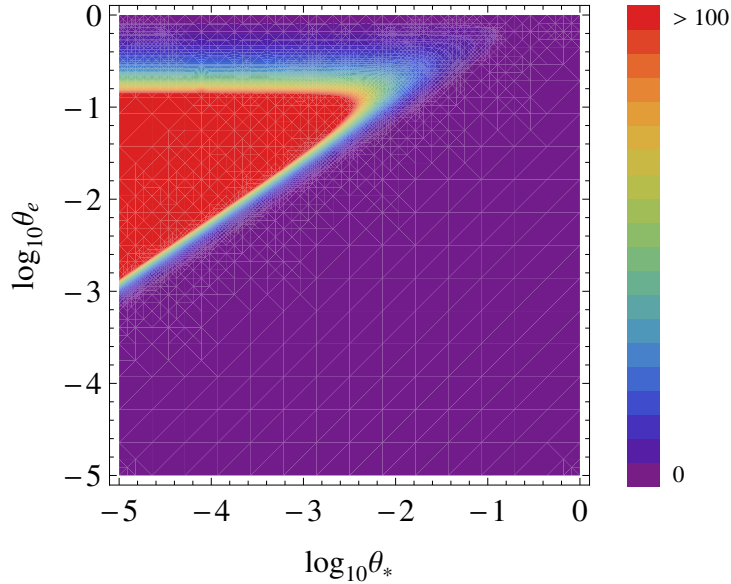}
 \end{center}
 \vspace{-2em}
 \caption{The prefactor of the last term of (\ref{nG}) as a function of $\theta_\star$ and $\theta_e$. There is a region where this factor blows up so that we can obtain a detectable value of $|\fnl |$. Note that we may have even larger regions that allow detectable $\fnl$ for $X>0$, which however means $\varphi_\star < \varphi_e$ so we do not consider this case.}
 \label{fig:fNLfactor}
\end{figure}

In Figure~\ref{fig:fNLfactor}, we show the prefactor of the last term of (\ref{nG}) as a function of $\theta_\star$ and $\theta_e$ with $X$ fixed by (\ref{b_difference}). The condition for large $\fnl$ is clear from this plot: the contribution of $\epsilon^\chi$ to the slow-roll condition should be larger at the end of inflation than at the moment of horizon crossing to occupy $10^{-4}\%\sim10\%$ of $\epsilon_e$. This means along the $\chi$ direction, the situation is more or less similar to the hilltop inflation, where the initial condition of placing $\chi_\star$ close to the local maximum of its cosine potential is not as severe as it sounds~\cite{hilltop}.

We can estimate the value of $\fnl$ as follows. For $\theta_e\ll1$, and for $\calA^2\tan^2\theta_\star \gtrsim \tan^4\theta_e$ and $\tan^2\theta_\star \lesssim \tan^2\theta_e$, from the $\eta^{\chi\chi}$ terms of (\ref{nG}), we find
\begin{equation} \label{fnlestimate}
\fnl \approx \frac{5}{6} \left( -e^{-X}\frac{\tan^6\theta_e}{{\cal A}^4\tan^4\theta_\star}\,\eta^{\chi\chi}_\star+2\sin^2\theta_e \frac{\tan^4\theta_e}{\calA^4\tan^4\theta_\star} \,\eta^{\chi\chi}_e \right) \, .
\end{equation}
From (\ref{epetachi}), in the region with a negative $\eta^{\chi\chi}$, using $e^{-2\varphi_*/\sqrt{6}}\ll 1$ and $e^{-2\varphi_e/\sqrt{6}} \sim 1/2$, we can replace $\eta^{\chi\chi}$'s by
\begin{align}
\eta^{\chi\chi}_\star \approx & -4\xi|\delta|\sqrt{1 - \epsilon_\star \, \frac{\sin^2\theta_\star}{2\xi\delta^2}} \, ,
\\
\eta^{\chi\chi}_e \approx & -8\xi|\delta| \sqrt{1 - \frac{\sin^2\theta_e}{4\xi\delta^2}} \, .
\end{align}
It is also possible that $\eta^{\chi\chi}_e$ is positive for $\xi|\delta|\gtrsim 0.1$ from (\ref{chievolution}) for $\chi_*$ being close to a local maximum. But, in this case, $\eta^{\chi\chi}_*\sim -0.1$, so the spectral index would be modified too much by $\chi$ from (\ref{spectralindex}). Therefore, for a typical value of $|\eta^{\chi\chi}|\sim 0.01$ during inflation, we restrict ourselves to the region with $\eta^{\chi\chi}_e<0$. Note that there are inequalities between $\theta_\star$, $\theta_e$ and $\xi\delta^2$ as
\begin{align}
\sin^2\theta_\star < & \frac{2\xi\delta^2}{\epsilon_\star} \, ,
\\
\sin^2\theta_e < & 4\xi\delta^2 \, .
\end{align}
Thus, by choosing appropriate values of $\theta_\star$ and $\xi$ in (\ref{fnlestimate}) and taking into account the evolution of the $\chi$ field from (\ref{chievolution}) for the corresponding value of $\theta_e$, we can have large $\fnl$ with observationally consistent $n_\zeta$. In Table~\ref{table:numbers} we show two sets of sample values. Note that $\fnl<0$ in models where non-Gaussianity is generated by a divergent trajectory falling of a ridge~\cite{Byrnes:2008wi,Elliston:2011dr}.

\begin{table}[t]
 \begin{center}
\begin{tabular}{ccccccc}
 \hline\hline
 $\xi$ & $\delta$  & $\theta_\star$ & $\theta_e$  & $n_\zeta$ & $\fnl$ & $\fnl^\text{(analytic)}$
 \\
 \hline
 10    & 0.00340 & $10^{-4}$      & $10^{-2.17}$ & $0.952$   & $-14.6$  & $-16.5$
 \\
 1     & 0.0380   & $10^{-5}$      & $10^{-2.75}$ & $0.959$   & $-60.2$  & $-63.7$
 \\
 \hline\hline
\end{tabular}
 \end{center}
 \caption{$n_\zeta$ and $\fnl$ with given $\theta_\star$, $\theta_e$, $\xi$ and $\delta$. The value of $\calP_\zeta$ is set by the COBE normalization $2.5\times10^{-9}$. Note that for these values of $\theta_\star$, $\theta_e$ and $\xi$, we find the boost factor of the power spectrum 1.06 and 1.03 respectively. In the last column we show the analytic estimate of $\fnl$ given by (\ref{fnlestimate}).}
 \label{table:numbers}
\end{table}

\section{Conclusions}
\label{sec:conc}

To summarize, we have studied a simple multi-field inflation model by introducing a complex scalar field with non-minimal coupling to gravity. We have considered the general action for the complex scalar field with dimension-4 interactions in the potential. Assuming that the non-minimal coupling breaks the global $U(1)$ symmetry by a small amount, the Lagrangian has been brought to a simple form with a non-canonical kinetic term for the angular mode $\chi$ and a product form of the potential for $\chi$ and the modulus $\varphi$. Then, the potential of $\varphi$ becomes flat at large field values due to the non-minimal coupling as in Higgs inflation, while a small breaking of the global $U(1)$ symmetry in the potential terms makes the potential for $\chi$ flat as well. This allows slow-roll two-field inflation. Using the $\delta{N}$ formalism, we have computed the power spectrum $\calP_\zeta$, the spectral index $n_\zeta$ and the non-linear parameter $\fnl$. With the hilltop type initial condition of $\chi$, we have shown that $\fnl$ can be large enough to be detected in near future. The generality of the hilltop type inflation suggests that the initial conditions for large non-Gaussianity are not as finely tuned as previously believed, and the possibility of large enough $\fnl$ is rather high. We have also presented sample sets of the parameters which give $|\fnl | = \mathcal{O}(10)$ as well as the observationally consistent $n_\zeta$.

Our toy model with a complex scalar field can be easily embedded in well-known particle physics models such as the Standard Model with two Higgs doublets. When the global $U(1)$ symmetry adopted in this paper is identified with the Peccei-Quinn (PQ) symmetry,  in the limit of a small violation of the PQ symmetry the pseudo-scalar Higgs boson can obtain a flat potential such that it is an inflaton candidate. It will be interesting to see what the constraints on two-field inflation we have found in this paper imply for the search of the additional Higgs boson at the Large Hadron Collider. We leave this work in a future study.

\subsection*{Acknowledgements}

We thank Ki Young Choi for helpful comments.
JG is grateful to the Institute for Advanced Study at the Hong Kong University of Science and Technology for hospitality while this work was under progress.
This work is partially supported by Korean-CERN fellowship.

\end{document}